\documentclass[prl,twocolumn,showpacs,preprintnumbers,amsmath,amssymb]{revtex4}
\usepackage{graphicx}
\usepackage{dcolumn}
\usepackage{bm}
\begin{document}
\title{On the stability and the similarity of $N=82$ isotones}
\author{P. Arumugam} 
\author{S.K. Patra} 
\author{A. Abbas}
\affiliation{Institute of Physics, Sachivalaya Marg, Bhubaneswar -
751 005, India.}

\begin{abstract}
Here we study the stability and the similarity of all even $58\leq Z\leq
70$, $N=82$ isotones.  We confirm the two decades old study of one
of the authors (AA) regarding the extra-ordinary stability and
the similarity of these nuclei. We present here a new evidence which
shows very strongly the said magicity of those nuclei and as such
there exists a new ``plateau'' of magicity. Three well studied
theoretical models which have been successfully applied in nuclear
physics are used here to study the above phenomena.  None of these
model is able to reproduce the similarity and the doubly magic
character of these nuclei.  Therefore this hints at ``new
physics'' in these $N=82$ isotones.

\end{abstract}

 \maketitle

The $N=82$ isotones display a high level of regularity and thus
have been of great interest to both the experimentalists and the theorists
\cite{Aba84,Ham90,And90,Hol97,Mat00}.  Though there have been
claims that $Z=64$ Gd isotone was something special, through
systematic study of diverse empirical data, it was argued by one of
the authors (AA) that there was a whole ``plateau'' of stability
for all even $58\leq Z\leq 70$, $N=82$ isotones and that as such
all these be treated as doubly magic nuclei \cite{Aba84}.  Also
all these nuclei were found to be strikingly similar to each other
\cite{Aba84}. The idea of ``changing magicities'' was rather novel
then. However today, one does speak of changing magicity for low
mass nuclei, for example new magic numbers at $N=$ 6, 14, 34 etc
\cite{Tan01,Thi00,Dlo03}. In this letter we would like to
reexamine this two decades old idea of Abbas \cite{Aba84}
regarding the ``plateau of doubly magic $N=82$ isotones and their
similarity. In particular we shall study a new ``smoking gun''
kind of evidence to confirm Abbas' plateau of stability
\cite{Aba84}. We shall also study this ``plateau'' of stability
using theoretical models like Skyrme--Hartree--Fock-Bogoliubov (HFB), 
relativistic mean field (RMF) theory etc.  We find that
all these models fail to reproduce the new magicities. 

Abbas has studied even $58\leq Z\leq 70$, $N=82$ isotones through
empirical evidence available then (i.e. 1983) \cite{Aba84}.  We
have reexamined the same in the light of more accurate and recent
experimentally available data.  We reconfirm the analysis as per
the empirical data.  We would not like to repeat the same here and
refer the reader to Ref. \cite{Aba84} as to the similarity and the magicity of
these nuclei.  However for the sake of
completeness we would like to summarize and tabulate a few of the
arguments therein \cite{Aba84}.
\begin{enumerate}
\item[(a)] In $Z=$ 58, 60, 62, 64, 66, 68, 70, $N=82$ isotones the
lowest first excited state (level $2^+$ or $3^-$) is very high
and steady at about 1.6 MeV for all the nuclei
\item[(b)] A study of low lying states in $N=81$, 80, 79 and 78
isotones with $Z=$ 58, 60, 62, 64, 66, 68, 70 indicates amazing
similarity and stability
\item[(c)] The ratio $E_1(4^+)/E_1(2^+)$ for all these nuclei is
$\sim$ 1.3.  This puts all these nuclei in the category of
``doubly-magic'' as per standard interpretation.
\item[(d)] The rms radii change very slowly as mass number changes over this
whole region. Plus many other arguments in support of the said statement regarding
the similarity and the stability of these nuclei.  Here
we discuss a further ``smoking gun'' kind of evidence in support
of the above idea.
\end{enumerate}

To discuss magicity one normally plots separation energies
$S_{1n}$  and $S_{2n}$ as a function of neutron number $N$ for a
particular proton number $Z$ or plot $S_{1p}$  and $S_{2p}$ as a
function of proton number $Z$ for a particular $N$.  If we do so
for $N=82$ isotones then we would like to point out here that the evidences
for magicity continue to be there but that it is not very
prominent\cite{Aba84}.  It is very common to plot $S_{2p}$ and look for kinks
as evidence for magicity \cite{Thi00,Dlo03,Tho03}.  However it
should be pointed out that this one in itself should not be taken as
very conclusive evidence in favour or against magicity \cite{Aba84,Plo81}.

We therefore plot separation energies differently.  We plot
$S_{1n}$ as a function of proton number $Z$ for a particular $N$.
We show this in Fig. 1 for the magic numbers $N=28$ and $N=50$ isotones.  We
immediately note that in $N=28$ case the magic number $Z=20$ and
$Z=28$ show up very prominently.  For $N=50$ case though the
$Z$-number does not pass through any standard magic number it does
indicate extra stability at $Z=34$, 36 and 38.  We know that the $Z=38$ case $^{88}$Sr
anyway is known to form a stable structure, enabling good shell
model description of $^{90}$Zr excited states \cite{Aba84}.  These
$N=50$ isotones actually are precursors of a more interesting
effect to be discussed below.

\begin{figure}
\includegraphics[width=0.95\columnwidth, clip=true]{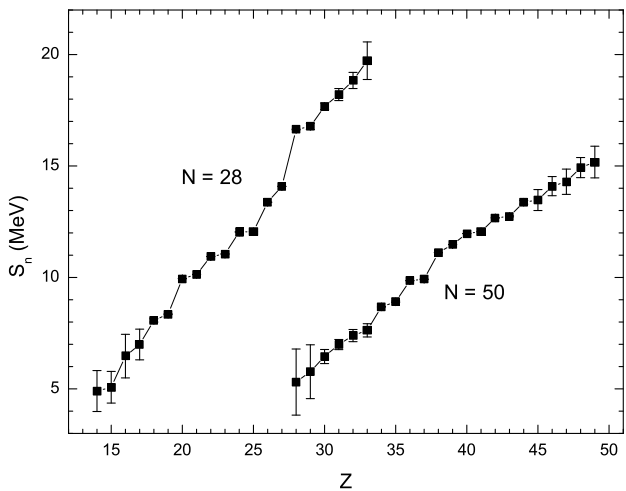}
\caption{Experimental single-neutron separation energies for
$N=28$ and $N=50$ isotones.}
\end{figure}

\begin{figure}
\includegraphics[width=0.95\columnwidth, clip=true]{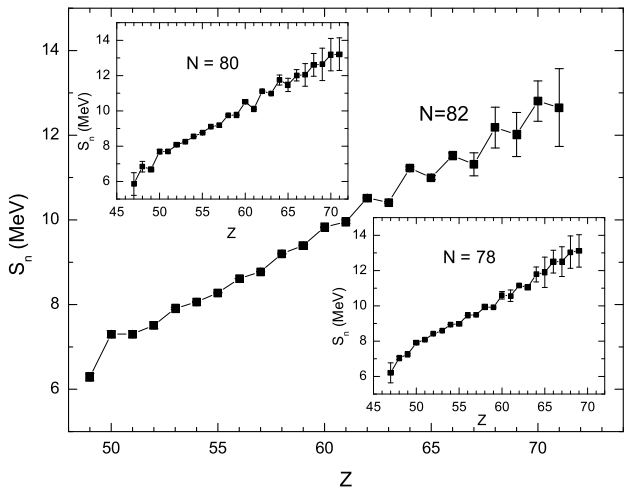}
\caption{Experimental single-neutron separation energies for
$N=82,\ 80$ and 78 isotones.}
\end{figure}

We plot $S_{1n}$ as a function of $Z$ for $N=82$ in Fig. 2.  The
magicity at $Z=50$ is very clearly demonstrated here.  But
amazingly exactly the same magicity is indicated, very prominently
at $Z=62$, 64, 68 and 70.  The magnitude of $S_{1n}$ at all these
neighbouring even $Z$-numbers is comparable to the magnitude of
$S_{1n}$ at $Z=50$.  We know $Z=50$ (Sn nucleus) is one of the ``best
'' magic number in nuclear physics.  As per Fig. 2 if we can call
$Z=50$ magic there is no reason why we should not do the same for
all the even $Z$ number discussed above.  This ``smoking gun''
evidence supplemented with those already provided by Abbas
\cite{Aba84} should convince one that all these nuclei are indeed
``doubly magic''.

\begin{figure}[t]
\includegraphics[width=1.1\columnwidth, clip=true]{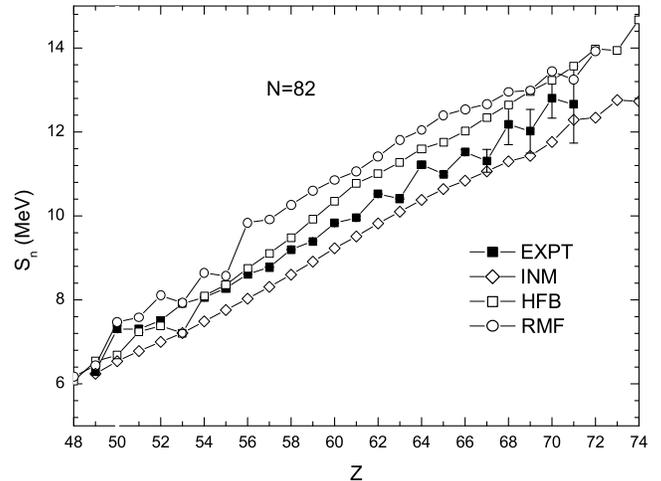}
\caption{Experimental single-neutron separation energies for
$N=82$ isotones compared with various theoretical model
calculations.}
\end{figure}

Does this magicity persist when two or even four neutrons are pulled out ?  The
corresponding $N=80$ plot (inset, Fig. 2) shows that indeed it is
so. We also plot the same for $N=78$ (inset, Fig. 2) which also
continues to show stability.  All this
should be treated as strong evidence for double magicity
of all these even $Z$ nuclei.  We would like to point out that we
do see similar effect for $N=50$, $N=48$ case but some other
evidences of double magicity and similarity are missing in these
nuclei (as discussed in ref. \cite{Aba84}) and hence these should
be actually treated as precursor of the amazing phenomenon pointed
out in this paper.

Today we do have elaborate theoretical framework of
various kind which are being successfully applied to study nuclear
phenomena.  We pick up three of these very successful models which are used
in nuclear physics.

We have carried out extensive study
microscopically in the nonlinear relativistic mean field theory of
Boguta and Bodmer \cite{12},
is an extended version of Walecka \cite{4} theory. We have adopted the NL3 \cite{14}
interactions in our study. The NL3 interaction has been widely used in recent
years in the calculation of varieties of nuclear properties like
binding energy, rms radii and giant resonances etc. and have been
accepted to be very successful. In the present study, we expanded
the fields in harmonic oscillator basis and studied the
stability of the result for each nucleus by varying the number of harmonic oscillator
shells between $N_F=N_B=12$ to 14.
We did the same exercise taking quadrupole deformation in our calculation, where the
basis deformation parameter $\beta_0$, was varied between $-0.4$ to 0.4
in the step of 0.1.
We used a constant gap BCS pairing calculation to
take into account the pairing correlation. The pairing constant
gap is taken for the drip-line nuclei following the prescription
of Medland and Nix \cite{16}. The formalism and calculation are quite
standard and have been widely used in the literature, the details
of which can be seen in Refs. \cite{17,18}. It is to be noted that
in the present study, we have performed three different calculations as to
pairing, i.e. (a) taking pairing in both even and odd
nucleon, (b) without pairing and (c) pairing correlation
is considered for even nucleons and neglected for the odd case (mixed pairing).
We found almost similar results for the without and the mixed pairing cases (cases (b) and
(c)). However, we noticed only a smooth increasing in $S_{1n}$ value for the case (a).
In this case, the sudden rise of $S_{1n}$ value at Z=50 could not be
reproduced (not shown in Fig. 3), which anyway is experimentally observed. On the other
hand for the other two
cases ((a) and (b)), the characteristic jump at Z=50 is clearly visible (here we 
have plotted
case (c) only in Fig. 3).

To see the other theoretical behaviour we have also displayed the calculated data of
HFB \cite{Sam01} and the 
infinite nuclear mass (INM) model
\cite{Sat01} calculations in
Fig. 3. From the figure it is clear that the HFB and INM 
models are not even able to reproduce the magic jump at Z=50
for the N=82 isotonic series. A further inspection of the figure, makes
it clear that the RMF formalism is somewhat able to reproduce the
known jump of Z=50, whereas it fails to reproduce the odd-even type
of staggering for $Z=$ 58, 60, 62, 64, 66, 68, 70, $N=82$ which are
experimentally observed. However this model produces some spurious 
tendency like a sudden jump
at $Z=56$ for the N=82 nucleus, which is experimentally ruled out and 
some odd-even type of staggering in the heavier region of the N=82
isotonic series.

For new magicities at $N=$ 6, 14, 16, 34 etc, all kind of new
ideas are being proposed \cite{Thi00,Dlo03,Tho03}.  We also feel
that the new magicities at $Z=58$, 60, 62, 64, 66, 68, 70 are also
indicative of new physics.  From our present investigation, it 
may be seen that the HFB and INM models are missing some important
physics to incorporate the experimental strggering for the 
considered region. On the other hand, the RMF explains the characteristic
jump at Z=50, but fails to reproduce other odd-even effects. The reproduction
of odd-even trends for heavier mass nuclei of the N=82 isotonic series in
the frame work of RMF model, gives some hints that the progress of
relativistic mean field formalism may be in the proper direction. However,
it is still missing some important ingredients. As it has been argued
by several authors, the inclusion of self-coupling of scalar fields simulate
the effect of three-body forces. But still the absence of many-body correlations is
very much there in this theory. A possible improvement of the RMF theory may be to 
include the higher order couplings as it is
suggested in Ref. \cite{aru03}. The other possible feature, which is
not taken either in the RMF or in the HFB or the INM models is the possibility 
of $A=3, 4 $ clustering \cite{abbas01}. Once these two effects are taken into
account, we hope that the theories may be compartible with the experimental
data. Work in this direction is in progress \cite{aru04}.

\end{document}